\documentclass[a4paper]{jpconf}

\usepackage{graphicx}
\usepackage{citesort}

\def\KogaKawakami{PRL.84.4461}

\def\Miyahara54{JPCM.15.327}
\def\ShastrySutherland{PB.108.1069} 
 
\def\CaV4O9{PRL.76.1932,PRL.76.3822} 
\def\Kageyama{PRL.84.5876}

\def\RestaReview{RestaReview}

\def\d#1{\,{\rm d}#1}

\def\vec#1{{\overrightarrow{#1}}}

\def\and{\,\,\&\,}

\def\ket#1{|#1\rangle }
\def\bra#1{\langle #1|}

\def\i{{\bf i}}

\def\skipm#1{}

\def\refsec#1{\S \ref{#1}}

\def\paragraph#1{}
\def\vec#1{{\mathbf{#1}}}
\def\i{i}

\begin{document}
\title{Z$_2$ topological number of local quantum clusters in the orthogonal dimer model}

\author{Isao Maruyama$^1$, Sho Tanaya$^2$, Mitsuhiro Arikawa$^{3,4}$, Yasuhiro Hatsugai$^2$}

\address{$^1$ Graduate School of Engineering Science, Osaka University, 
  Toyonaka, Osaka 560-8531, Japan}
\address{$^2$ Institute of Physics, Univ. of Tsukuba
  1-1-1 Tennodai, Tsukuba Ibaraki 305-8571, Japan}
\address{$^3$ Center for Computational Sciences, Univ. of Tsukuba
  1-1-1 Tennodai, Tsukuba Ibaraki 305-8571, Japan}
\address{$^4$ CREST, Japan Science and Technology Agency, Chiyoda-ku, Tokyo, 102-0075, Japan}

\ead{maru@mp.es.osaka-u.ac.jp}

\begin{abstract}
  We have studied the $Z_2$ topological number defined by the Berry phase
  for the gapped frustrated systems including the orthogonal dimer model which has
  a direct product state of local quantum clusters as the exact ground state.
  The $Z_2$ topological number can clarify what kind of the local quantum clusters is formed
  to lift the macroscopic degeneracy due to frustration,
  even when the exact ground state is unknown.
  As a demonstration,
  the dimer-singlet  and plaquette-singlet phase
  are identified by two kinds of Z$_2$ topological numbers
  in the Shastry-Sutherland model
  and its generalization
  realized  experimentally
  as SrCu$_2$(BO$_3$)$_2$ and CaV$_4$O$_9$.
\end{abstract}

\section{Introduction}
Macroscopic degeneracy that survives at low temperature is a key
feature of frustrated materials.  Approaching the absolute zero
temperature, the degeneracy will be lifted to meet the third law of
thermodynamics. How is the degeneracy resolved?  Its answer has a rich
diversity, because frustrated systems have many exotic phases that are
sensitive to small change.  A standard scenario is the symmetry
breaking mechanism that tells us an order parameter to characterize
how the degeneracy is resolved.  However, recent development in
condensed matter physics reveals that there are topological phases
where usual order parameters do not play any fundamental roles.
Instead of usual order parameters such as a magnetization, topological
invariants are used to characterize the topological phases and they
are linked to the presence or absence of gapless edge modes.  A
historical example is the quantum Hall system where the topological
invariant is the Chern number that is an integer defined by the Berry connection
theoretically and observed as the Hall conductance experimentally.
In addition to the Chern number, Z$_2$ topological values, i.e., the
integers modulo two, have been focused in recent studies on quantum
spin Hall systems\cite{AX.1002.3895,NP.5.378}. In such topological
phases, many of them are gapped.

In some frustrated materials, gap formation is another possible
scenario for lifting the macroscopic degeneracy as in the
two-dimensional gapped spin system,
SrCu$_2$(BO$_3$)$_2$\cite{\Kageyama,JPSJ.79.011005}.  A discovery of
SrCu$_2$(BO$_3$)$_2$ shed a light on theoretical construction of the
model Hamiltonian with an exact ground state of local spin-singlets,
i.e., the Shastry-Sutherland
model\cite{\ShastrySutherland,PRL.82.3701}.  Inspired by Majumdar's
one-dimensional model\cite{JPC.12.735}, Shastry and Sutherland
constructed the two-dimensional geometrically frustrated spin system
in which the exact ground state is the dimer-singlet(DS) state
which is
a direct product state  of local spin-singlets\cite{\ShastrySutherland}.  Unless the
spin gap closes, the DS state is still the exact ground state against
the particular kind of perturbation due to the orthogonality of the
dimers.  Such construction is common in orthogonal dimer models
including three-dimensional generalization\cite{JPCM.15.327}.

The
Shastry-Sutherland model also has an gapped phase, the
plaquette-singlet(PS) phase\cite{\KogaKawakami,JPSJ.70.1369,PRB.66.014401}, 
in addition to the DS phase.
Unlike the DS phase, it is difficult to obtain its exact ground state in the PS phase.
In fact, Weihong {\it et.al.}\cite{\Miyahara54} have raised a possibility that the 
PS phase is unstable.
To clarify it, one can consider an adiabatic modification of the Hamiltonian connected to the decoupled
PS Hamiltonian which has a direct product state of PSs as the exact ground state.
Actually, the ground state in the PS phase is adiabatically connected to that of the decoupled PS Hamiltonian\cite{\KogaKawakami}.
The generalized theoretical model discussed in the adiabatic connection 
includes the 1/5-depleted square lattice that
is realized as CaV$_4$O$_9$\cite{\CaV4O9}.
In this sense,
the direct product state of local quantum objects, such as DSs and PSs,
and 
the adiabatic modification toward decoupled Hamiltonians are not just theoretical toys.

\paragraph{bulk-edge}
Recently, 
a method to obtain a decoupled Hamiltonian
is proposed,
that is, the $Z_2$ topological number defined by the Berry phase\cite{JPSJ.75.123601,PRB.78.054431,PRB.79.115107,PRB.79.205107,PRB.82.073105}.
Unless the spin gap closes,
the $Z_2$ topological number remains invariant, i.e., topologically protected against  perturbation.
These situations are quite similar to the case of
the Chern number.
The difference is that the Chern number is integer
while 
the $Z_2$ topological number is 
quantized to the following two values: a trivial value 0 or nontrivial value $\pi$ (mod $2\pi$)
due to the time reversal symmetry of the spin system.
However, since the Berry phase is generally defined through the local perturbation,
we can consider many kinds of the $Z_2$ topological numbers defined by the Berry phase.
The $Z_2$ topological numbers have successfully identified
the spin-singlets,  plaquette-singlets, and Haldane state in the spin ladders.
It can be applied to not only the spin system but also the electron system including the BCS system\cite{PRB.82.073105}.
In addition, the itinerant singlets and the Kondo singlets are identified.
Recently, we have succeeded in generalizing the $Z_2$ Berry phase into $Z_Q$ Berry phase with any integer $Q$\cite{AX.1009.3792}.

\begin{figure}
  \resizebox{22pc}{!}{\includegraphics{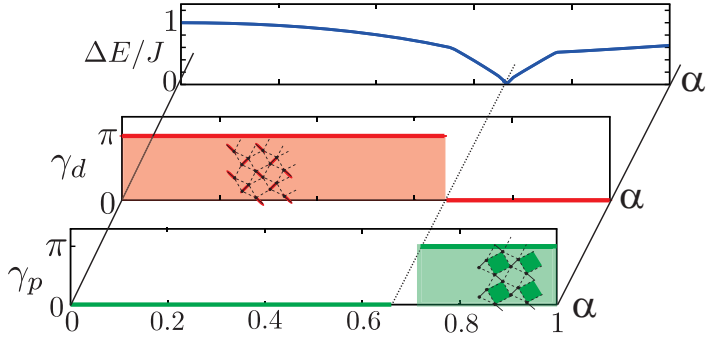}}\hspace{2pc}%
\begin{minipage}[b]{14pc}
  \caption{\label{fig:SS}Gap $\Delta E$ and Berry phases of the Shastry-Sutherland model
    for system size $N=16$ with the adiabatic parameter $\alpha=J'/J$.
    $J$ (and $J'$) is the nearest neighbor (next nearest neighbor) interaction.
    $\gamma_{\rm d}$ and $\gamma_{\rm p}$ are Berry phases for the DS and PS,
    respectively.
    The detail is discussed in \refsec{sec:SSmodel}.
  }\end{minipage}
\end{figure}

\paragraph{in this letter}
In this paper, we apply the $Z_2$ topological number to the Shastry-Sutherland model.
Our numerical result in Fig.~\ref{fig:SS}
shows that
the DS and PS phase are clearly identified by two kinds of $Z_2$ topological numbers, $\gamma_{\rm d}$ and $\gamma_{\rm p}$, respectively.
Numerical detail will be explained in \refsec{sec:SSmodel},
but here we note that the ground state in the PS phase is 
not a simple direct product state, while there is the simple exact ground state in the DS phase.
Generally speaking,
although
the direct product state
is not an exact ground state 
when small long range Heisenberg exchange interactions
are considered as a perturbation for real material of SrCu$_2$(BO$_3$)$_2$,
it will be  widely accepted that
the ground state in the perturbation
is similar to the direct product state.
The similarity is naively based on the adiabatic connection of two ground states.
By using the $Z_2$ topological classification,
it is possible to discuss  clear equivalence of ground states via the adiabatic connection,
as we will explain in \refsec{sec:def}.
This approach is demonstrated for the generalized theoretical model including SrCu$_2$(BO$_3$)$_2$ and  CaV$_4$O$_9$
in \refsec{sec:SSmodel}.
\section{Definition of the Z$_2$ topological number for orthogonal dimer models\label{sec:def}}
\paragraph{local cluster}
To describe our approach, let us start with local quantum clusters.
A local quantum state $\ket{\psi_i}$ is supposed to be the singlet ground state of a quantum local Hamiltonian $h_i$ of 
an $i$th cluster,
such as a spin-singlet, plaquette-singlet, local Kondo-singlet 
and dimerized electrons state.
The decoupled Hamiltonian is given by
\begin{math}
  H_0 = \sum_i h_i   
\end{math}
with
$[h_i,h_j]=0$.
Due to small system-size of $h_i$,
the energy gap $\Delta$ generally opens
above the exact ground state $\ket{\Psi_0} = \prod_i \ket{\psi_i}$.
Although the local singlet-cluster $\ket{\psi_i}$ 
is a pure quantum object and does not have a classical analogue,
the direct product state $\ket{\Psi_0}$
is a classical state in the sense that there is no 
quantum entanglement
between the clusters.
The decoupled Hamiltonian is a start point to consider both construction of the orthogonal dimer model
and definition of the $Z_2$ topological number.

To construct an orthogonal dimer model generically,
we connect the decoupled Hamiltonian $H_0$ with an interaction $H_1$.
Introducing an adiabatic parameter $\alpha$,
we define $H_\alpha = (1-\alpha)H_0 +  \alpha H_1$ or $H_\alpha = H_0 + \alpha H_1$.
To keep the direct product state $\ket{\Psi_0}$ as the exact ground state for $\alpha>0$,
a condition $H_1 \ket{\Psi_0} =0$ is required.
Since we supposed $\ket{\psi_i}$ is a singlet-state, 
$H_1$ can be constructed by a total spin operator for $i$th cluster
$T_{i}^{\beta}=\sum_{l} S^\beta_{il}$ ($\beta=x,y,z$) with 
$T_{i}^{\beta} \ket{\psi_{i}} = 0$,
where $il$ indicates the $l$th site in the $i$th cluster.
The condition guarantees that the ground state of $H_\alpha$ 
is the direct product state around $\alpha=0$ to some extent as far as the gap is non-zero.

\paragraph{construction of Shastry-Sutherland}
In the orthogonal dimer model,
we set 
the local spin 1/2 cluster 
$h_i = J \vec{S}_{i1}\cdot \vec{S}_{i2}$,
with the local singlet ground state $\ket{\psi_{i}}={1\over \sqrt{2}} \left(\ket{\uparrow}_{i1}\ket{\downarrow}_{i2}
  -\ket{\downarrow}_{i1}\ket{\uparrow}_{i2}\right)$.
Using the total spin operator $T_{i}^\beta = S^\beta_{i1}+ S^\beta_{i2}$,
we can construct 
\begin{math}
  H_1 =  \sum_{i,i'\neq i,l'} J'_{ii'l'} \vec{S}_{i'l'} \cdot (\vec{S}_{i1} + \vec{S}_{i2})
  .
\end{math}
The coefficients $J'_{ii'l'}$ are determined by the lattice geometry.
In this construction, the orthogonality of the dimer bonds 
is always existing, that is, the interaction between $i1$ and $i'l'$ is identical to that between $i2$ and $i'l'$ always.
$H_1$ can be generalized to a scalar product type
$H_1 = \sum J'_{ij\beta j'\beta'} \left( \vec{S}_{j\beta} \times \vec{S}_{j'\beta'}\right) \cdot (\vec{S}_{i1}+ \vec{S}_{i2})$
and more complicated types.
In addition, 
the local quantum objects can be generalized
to a plaquette singlet when we consider 
$h_i = J (
\vec{S}_{i1}\cdot \vec{S}_{i2}
+\vec{S}_{i2}\cdot \vec{S}_{i3}
+\vec{S}_{i3}\cdot \vec{S}_{i4}
+\vec{S}_{i4}\cdot \vec{S}_{i1}
)$
with $T_{i}^\beta= \sum_{l=1}^4 S^\beta_{il} = {T_{i}^\beta}^\dagger$.
In the same manner, one can consider larger cluster singlets
and generalizations to local quantum objects
of bosonic or fermionic systems.
Note that the Hermite condition $H_1 = H_1^\dagger$ imposes some limitation on $H_1$.
In the above construction,
the gap $\Delta$ of $H_0$, which comes from a finite-size gap of the local quantum cluster, plays an important role for the stability of the exact ground state against the adiabatic parameter $\alpha$.

If the local ground state $\ket{\psi_{i}}$ is not a singlet but a multiplet, 
we must deal with the macroscopic degeneracy.
The gap will open above the macroscopically degenerate states
and is identified by the $Z_2$ topological number defined by non-Abelian Berry phases.
Of-course, the perturbation $H_1$
may lift the degeneracy and we need further discussion if we focus on the states below newly opened gap.
In the following, we limit ourselves to a non-degenerate ground state and Abelian Berry phases in the definition of the $Z_2$ topological number.

\paragraph{Berry's phase}
The Berry's geometrical phases\cite{\RestaReview} are used to calculate the macroscopic
polarization in the solid with surfaces.  We emphasize that an edge or surface
property such as the macroscopic polarization can be obtained via the Berry phase which is calculated in the
periodic bulk-system without surfaces.  This is bulk-edge
correspondence, which is originally proposed as the
one-to-one correspondence between the Chern number defined by the
Berry connections and the number of the
edge states  in the quantum Hall systems\cite{PRL.71.3697}.  Such local degrees of freedom on the boundaries are also
characteristic in the topological insulators, and quantum spin Hall
systems\cite{AX.1002.3895,NP.5.378}.
The bulk-edge correspondence is important
for the $Z_2$ topological number used in this paper.
For example, the non-trivial $Z_2$ topological number guarantees the existence
of the Kennedy triplet for an open chain\cite{PRB.79.205107}.

To illustrate the definition of the $Z_2$ topological number for $H_\alpha$, let us consider the decoupled Hamiltonian $H_0$ first.
In our previous studies,
it have been clarified that
the gapped local quantum cluster,
such as 
the spin-singlet, plaquette-singlet, and Kondo singlet, gives the non-trivial $\pi$ Berry phase.
The key ingredient to obtain the $\pi$ Berry phase for the local cluster $h_i$
is the unitary gauge transformation.
Let us explain it with detailed definition of the Berry phase\cite{PRL.71.3697}.
The Abelian Berry phase is defined by a local $SU(2)$ spin twist on a specified site.
When we choose a local cluster $h_I$ and a $l$th site in $h_I$, 
the local spin twist is introduced by
the substitution $h_I \rightarrow h_I(\phi)=U^\dagger(\phi) h_I U(\phi)$ 
with $U(\phi)=\exp\left[ \i \phi (S-S_{Il}^z) \right]$ for a spin $S=1/2$,
where $Il$ indicates the chosen site.
With using $h_I(\phi)$,
the decoupled Hamiltonian has one-parameter dependence 
as $H_0(\phi)= h_I(\phi) + \sum_{i'\neq I} h_{i'}$.
With the adiabatic parameter $\alpha$
we define $H_\alpha(\phi) = (1-\alpha) H_0(\phi) + \alpha H_1$
or $H_\alpha(\phi) = H_0(\phi) + \alpha H_1$.

For an one-parameter dependent Hamiltonian $H_\alpha(\phi)$,
the Berry phase $\gamma$ is defined as
\begin{math}
  \gamma = \int_0^{2\pi} \d{\phi} \bra{\Psi_\alpha(\phi)}{\d{}\over \d{\phi}}
  \ket{\Psi_\alpha(\phi)}
  ,
\end{math}
where $\ket{\Psi_\alpha(\phi)}$ is the ground state of $H_\alpha(\phi)$.
The numerical integration 
is given
by discretizing the parameter space of 
$\phi$ into finite number of points\cite{PRB.47.1651},
which is enough up to 60.
The Berry phase here is quantized into $0$, or $\pi$ if the ground state is
invariant under an anti-unitary operator.
The anti-unitary operator is 
the time-reversal operator for the spin system.
For the decoupled Hamiltonian $H_0(\phi)$,
the ground state is a direct product state
and has one-parameter dependence through the unitary transformation $\ket{\Psi_0(\phi)} = U(\phi)\ket{\Psi_0(0)}$,
which gives the $\pi$ Berry phase,
because
$\gamma= 2\pi (S-\bra{\Psi(0)}S_{Il}^z \ket{\Psi(0)})=\pi$ in the spin $S=1/2$ case.
This result does not depend on
the detail of the local Hamiltonian $h_i$.
When the Hamiltonian is $H_\alpha(\phi) = (1-\alpha) H_0(\phi) + \alpha H_1$,
there should be the change of the $Z_2$ topological number.
At $\alpha=1$,
since the Hamiltonian $H_{\alpha=1}(\phi)=H_1$ does not depend on $\phi$,
the Berry phase $\gamma$ cannot become the non-trivial value $\pi$,
that is,
$\gamma$ is zero if the gap opens or undefined if the gap closes.
Then, there is inevitably the change of $\gamma$ in the adiabatic modification via $\alpha$,
which signals whether the ground state is equivalent to the direct product states 
or not.
It should be noted that since $H_\alpha(\phi) \neq U^\dagger(\phi) H_\alpha U(\phi)$ due to $H_1$,
the Hamiltonian $H_\alpha(\phi)$ is able to have a local flux, which cannot be gauged out.
Then, the problem is non-trivial for generic $H_1$.

\begin{figure}
  \resizebox{23pc}{8pc}{\includegraphics{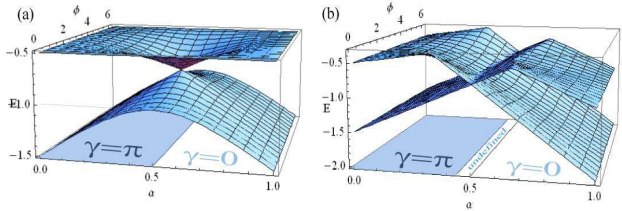}}
  \hspace{2pc}%
  \begin{minipage}[b]{12pc}
  \caption{  \label{fig:PSDS}Two typical energy diagrams as a function of $\phi$ and $\alpha$
    of $H_\alpha(\phi) = (1-\alpha) H_0(\phi) + \alpha H_1$
    with (a) the non-orthogonal interaction $H_1= \vec{S}_{12}\cdot \vec{S}_{21} + \vec{S}_{11}\cdot \vec{S}_{22}$
    and (b) the orthogonal interaction $H_1= (\vec{S}_{11}+ \vec{S}_{12})\cdot (\vec{S}_{21} + \vec{S}_{22})$.
  }
\end{minipage}
\end{figure}

\paragraph{PS and DSs}
Typically, there are two types of the transitions for the $Z_2$ Berry phase in the adiabatic modification $\alpha$.
To discuss it,
let us show an example of a local cluster, the $N=4$ spin $S=1/2$ system
with the decoupled Hamiltonian $H_0(\phi)=U^\dagger (\phi) h_1 U(\phi) + h_2$
of the dimer singlets $h_{i}=\vec{S}_{i1}\cdot \vec{S}_{i2}$
and the local spin twist $U(\phi)=\exp\left[ \i \phi (S-S_{11}^z) \right]$.
Here we consider two kinds of Hamiltonians $H_\alpha(\phi)=(1-\alpha)H_0(\phi)+\alpha H_1$:
(a)  the non-orthogonal interaction $H_1= \vec{S}_{12}\cdot \vec{S}_{21} + \vec{S}_{11}\cdot \vec{S}_{22}$
and (b) the orthogonal interaction $H_1= (\vec{S}_{11}+ \vec{S}_{12})\cdot (\vec{S}_{21} + \vec{S}_{22})$.
Due to the lattice symmetry, the Berry phase does not depend on the site where $U(\phi)$ is introduced.
In the case (b), 
the exact ground state of $H_\alpha(0)$ is a direct product state of the dimer singlets for $\alpha<\alpha_c$
and there is a level cross at $\alpha=\alpha_c=1/2$.
On the contrary, 
in the case (a),
the direct product state is not an exact ground state for $\alpha>0$
and
there is finite gap without
any level cross in the range of $\alpha \in [0,1]$ for $H_\alpha(0)$.
We expect that in the latter adiabatic modification 
two DSs delocalize asymmetrically 
except for the symmetric point $\alpha=\alpha_c$.
The $Z_2$ Berry phase detects the asymmetry of the DSs.
In Fig.~\ref{fig:PSDS},
the result of the $Z_2$ Berry phase is shown in both cases (a) and (b).
In the case (a),
the $Z_2$ Berry phase clearly identified the two phases connected to the two decoupled models, $\alpha=0$ and $1$, respectively.
At the phase boundary $\alpha=\alpha_c$, the  $Z_2$ Berry phase becomes undefined
because the level cross occurs at $\phi=\pi$.
In other words,
the diabolic point of the Dirac cone generates the $\pi$ Berry phase.
Here, we emphasize that the Hamiltonian $H_\alpha(0)$ at $\phi=0$ has no gap closing 
in the adiabatic modification via $\alpha$
and there is not a transition but a cross-over\cite{PRB.82.073105}.
In the case (b) as shown in Fig.~\ref{fig:PSDS}(b),
the level cross at $\alpha=\alpha_c$ without twist $\phi$
induces a finite range of the undefined region of the $Z_2$ Berry phase.

\section{Results on the Shastry-Sutherland model\label{sec:SSmodel}}
\begin{figure}
\begin{minipage}[b]{10pc}
  \centering
  \resizebox{3cm}{!}{\includegraphics{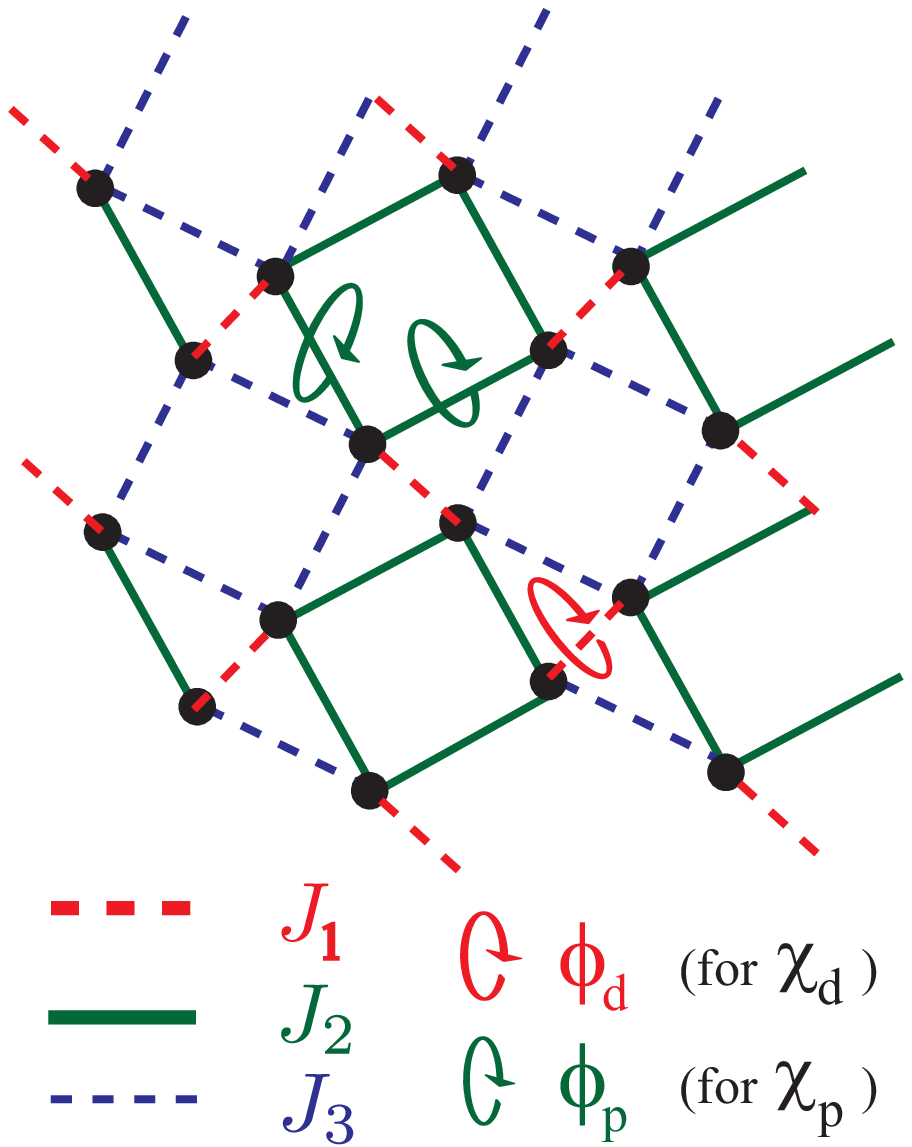}}
  \caption{\label{fig:model} Hamiltonian with $J_1,J_2,$ and $J_2$.
  }
\end{minipage}
\hspace{2pc}
\begin{minipage}[b]{26pc}
\centering
  \resizebox{7cm}{!}{\includegraphics{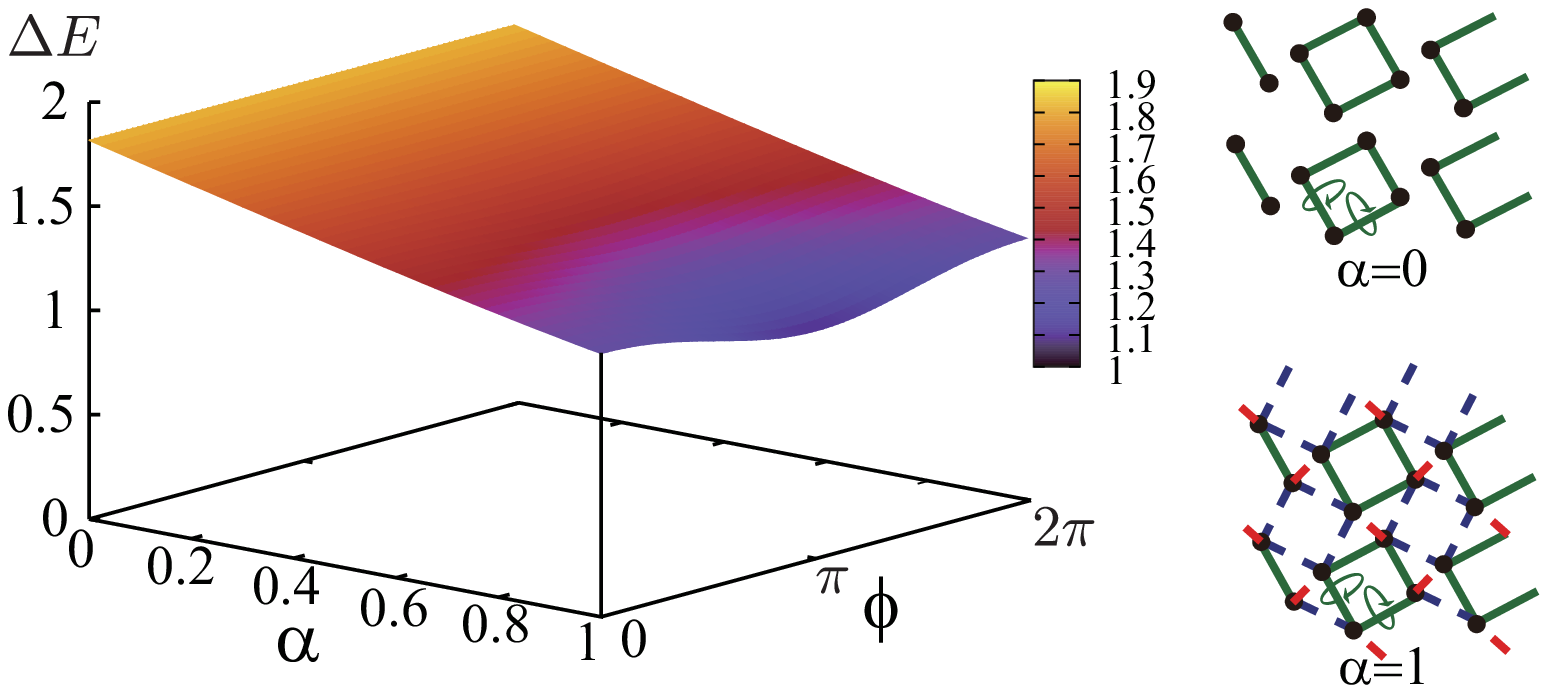}}
  \caption{  \label{fig:PS}The gap $\Delta E$ for the $Z_2$ topological number $\gamma_{\rm p}$.
    The twist $\phi=\phi_{\rm p}$ is defined for the decoupled PS Hamiltonian with the bonds $J_2=1.82$.  
    The adiabatic parameter $\alpha$ is introduced as $J_1 = \alpha, J_3=0.95 \alpha$.
  }
\end{minipage}
\end{figure}

\paragraph{Shastry-Sutherland}
Here we discuss adiabatic modifications of the Shastry-Sutherland model
as shown in Fig.~\ref{fig:SS}.
The Shastry-Sutherland model with $J$ and $J'=\alpha J$ can be defined as
the case $J_1=J$ and $J_2=J_3=J'$ of
the generalized model as shown in Fig.~\ref{fig:model}.
As mentioned in previous sections,
the definition of the Berry phase depends on the local spin twist
and on the choice of the decoupled Hamiltonian for a given Hamiltonian.
For this Shastry-Sutherland model, $\gamma_{\rm d}$ (and $\gamma_{\rm p}$) 
is defined by the local spin twist $\phi_{\rm d}$ ($\phi_{\rm p}$) shown in Fig.~\ref{fig:model}.
The decoupled DS (and PS) Hamiltonian only with $J_1$ ($J_2$) corresponds to $\gamma_{\rm d}$ ($\gamma_{\rm p}$).
Since the Shastry-Sutherland model has the exact ground state of the DSs,
the level cross at $\alpha=\alpha_c$ is expected as in Fig.~\ref{fig:PSDS}(b).
Actually, in our calculation with the small system size $N=16$, 
there is the first-order quantum phase transition, that is, a gap closing $\Delta=0$ occurs
at $\alpha_{c}= 0.67$ as shown in Fig.~\ref{fig:SS}.
Moreover, $\gamma_{\rm d}$ shows a clear transition at $\alpha=\alpha_{c}$ in Fig.~\ref{fig:SS}.
This is quite natural because we know the exact DS ground state for $\alpha=J'/J<\alpha_{c}$.
The question arising here is how the ground state for $\alpha> \alpha_{c}$
is identified.
To answer this question we introduce another $Z_2$ topological number, $\gamma_{\rm p}$,
which corresponds to the decoupled PS Hamiltonian 
having the direct product state of local PSs as the exact ground state.
Figure~\ref{fig:SS} shows that the two $Z_2$ topological numbers
can identify the DS phase for $\alpha<\alpha_{c}$ and the PS phase for $\alpha>\alpha_{c}$.
However, since there is the first-order phase transition at $\alpha_c$,
the $Z_2$ Berry phase will show instability just around the level cross point as shown in Fig.~\ref{fig:PSDS}(b).
In fact, $\gamma_{\rm p}$
shows an undefined region around $\alpha=\alpha_{c}$ as shown in Fig.~\ref{fig:SS}.
Although the system size is too small and the transition point $\alpha_{c}$ is not comparable to previous studies,
the $Z_2$ Berry phase clarifies a PS-type character of the ground state for $\alpha>\alpha_{c}$.

In addition, we consider $J_2/J_1 = 1.82$ and $J_3/J_1 = 0.95$
in the parameter space $J_1,J_2,$ and $J_3$ of the generalized model in Fig.~\ref{fig:model},
which corresponds to CaV$_4$O$_9$ and is in the PS phase\cite{JPSJ.70.1369}.
To clarify the property of the ground state by the $Z_2$ topological number,
we calculated $\gamma_{\rm p}$ and we obtained  $\gamma_{\rm p}=\pi$ for $N=16$ sites.
To illustrate the result,
we show the gap $\Delta E$ as
a function of twist $\phi=\phi_{\rm p}$ and the adiabatic parameter $\alpha$
from the decoupled PS Hamiltonian at $\alpha=0$ to CaV$_4$O$_9$ at $\alpha=1$.
The decoupled PS Hamiltonian has $\gamma_{\rm p}=\pi$ by definition.
Figure \ref{fig:PS} shows that there is no gap closing in the adiabatic modification
not only at $\phi=0$ but in whole region of $\phi \in [0,2\pi]$.
To change the $Z_2$ topological number, the gap closing is required as discussed in \refsec{sec:def}.
In this sense, $\gamma_{\rm p}=\pi$ means that the state is adiabatically connected to
the direct product state in the two dimensional parameter space including the gauge twist angle $\phi \in [0,2\pi]$.

\section{Summary}
\paragraph{summary}
In summary, we have studied the $Z_2$ topological number of the orthogonal dimer model as the gapped frustrated spin system.
The construction of the orthogonal dimer model with the exact ground state
and the $Z_2$ topological number $\gamma_{\rm d}$ are based on the same decoupled DS Hamiltonian.
As shown in Fig.~\ref{fig:SS},
the DS phase has the non-trivial $Z_2$ topological number, $\gamma_{\rm d}=\pi$,
corresponding the direct product state of DSs.
The orthogonal dimer model shows the first-order transition with a level crossing,
as shown in Fig.~\ref{fig:PSDS}(b),
in the finite-size calculation.
Although the exact ground state after the transition is unknown,
the $Z_2$ topological number which corresponds to the decoupled PS Hamiltonian
shows $\gamma_{\rm p}=\pi$
for $\alpha>\alpha_{c}$ as shown in  Fig.~\ref{fig:SS}
and for the specified parameter corresponding to CaV$_4$O$_9$ as clarified in Fig.~\ref{fig:PS}.
Even though our analysis using the exact diagonalization is limited for the small system-size,
this result agrees with existence of the PS phase discussed in the previous studies\cite{PRL.84.4461,JPSJ.70.1369}.

\ack
The authors are grateful to A. Koga for his valuable discussion.
This work was supported by a Grant-in-Aid from
the Ministry of Education, No. 20340098 from JSPS, No.22014002 on
Priority Areas from MEXT.
Some numerical
calculations were carried out on Altix3700BX2 at YITP in Kyoto
University
and the facilities of the Supercomputer Center, 
Institute for Solid State Physics, University of Tokyo.

\section*{References}
\providecommand{\newblock}{}

\end{document}